\documentclass[a4paper,11pt]{article}

\usepackage{jheppub} 

\usepackage{graphicx}
\usepackage{color}
\usepackage{mathtools}
\usepackage[left]{lineno}
\usepackage{subfig}

\title{Ortho-positronium observation in the Double Chooz Experiment}

\newcommand{\Aachen}{III. Physikalisches Institut, RWTH Aachen University, 52056 Aachen, Germany}
\newcommand{\Alabama}{Department of Physics and Astronomy, University of Alabama, Tuscaloosa, Alabama 35487, USA}
\newcommand{\Argonne}{Argonne National Laboratory, Argonne, Illinois 60439, USA}
\newcommand{\APC}{AstroParticule et Cosmologie, Universit\'{e} Paris Diderot, CNRS/IN2P3, CEA/IRFU, Observatoire de Paris, Sorbonne Paris Cit\'{e}, 75205 Paris Cedex 13, France}
\newcommand{\CBPF}{Centro Brasileiro de Pesquisas F\'{i}sicas, Rio de Janeiro, RJ, 22290-180, Brazil}
\newcommand{\Chicago}{The Enrico Fermi Institute, The University of Chicago, Chicago, Illinois 60637, USA}
\newcommand{\CIEMAT}{Centro de Investigaciones Energ\'{e}ticas, Medioambientales y Tecnol\'{o}gicas, CIEMAT, 28040, Madrid, Spain}
\newcommand{\Columbia}{Columbia University; New York, New York 10027, USA}
\newcommand{\Davis}{University of California, Davis, California 95616, USA}
\newcommand{\Drexel}{Department of Physics, Drexel University, Philadelphia, Pennsylvania 19104, USA}
\newcommand{\Hiroshima}{Hiroshima Institute of Technology, Hiroshima, 731-5193, Japan}
\newcommand{\IIT}{Department of Physics, Illinois Institute of Technology, Chicago, Illinois 60616, USA}
\newcommand{\INR}{Institute of Nuclear Research of the Russian Academy of Sciences, Moscow 117312, Russia}
\newcommand{\CEA}{Commissariat \`{a} l'Energie Atomique et aux Energies Alternatives, Centre de Saclay, IRFU, 91191 Gif-sur-Yvette, France}
\newcommand{\Kansas}{Department of Physics, Kansas State University, Manhattan, Kansas 66506, USA}
\newcommand{\Kobe}{Department of Physics, Kobe University, Kobe, 657-8501, Japan}
\newcommand{\Kurchatov}{NRC Kurchatov Institute, Moscow 123182, Russia}
\newcommand{\MIT}{Massachusetts Institute of Technology, Cambridge, Massachusetts 02139, USA}
\newcommand{\MaxPlanck}{Max-Planck-Institut f\"{u}r Kernphysik, 69117 Heidelberg, Germany}
\newcommand{\Niigata}{Department of Physics, Niigata University, Niigata, 950-2181, Japan}
\newcommand{\NotreDame}{University of Notre Dame, Notre Dame, Indiana 46556, USA}
\newcommand{\IPHC}{IPHC, Universit\'{e} de Strasbourg, CNRS/IN2P3, 67037 Strasbourg, France}
\newcommand{\SUBATECH}{SUBATECH, CNRS/IN2P3, Universit\'{e} de Nantes, Ecole des Mines de Nantes, 44307 Nantes, France}
\newcommand{\Tennessee}{Department of Physics and Astronomy, University of Tennessee, Knoxville, Tennessee 37996, USA}
\newcommand{\TohokuUni}{Research Center for Neutrino Science, Tohoku University, Sendai 980-8578, Japan}
\newcommand{\TohokuGakuin}{Tohoku Gakuin University, Sendai, 981-3193, Japan}
\newcommand{\TokyoInst}{Department of Physics, Tokyo Institute of Technology, Tokyo, 152-8551, Japan }
\newcommand{\TokyoMet}{Department of Physics, Tokyo Metropolitan University, Tokyo, 192-0397, Japan}
\newcommand{\Muenchen}{Physik Department, Technische Universit\"{a}t M\"{u}nchen, 85748 Garching, Germany}
\newcommand{\Tubingen}{Kepler Center for Astro and Particle Physics, Universit\"{a}t T\"{u}bingen, 72076 T\"{u}bingen, Germany}
\newcommand{\UFABC}{Universidade Federal do ABC, UFABC, Santo Andr\'{e}, SP, 09210-580, Brazil}
\newcommand{\UNICAMP}{Universidade Estadual de Campinas-UNICAMP, Campinas, SP, 13083-970, Brazil}
\newcommand{\vtech}{Center for Neutrino Physics, Virginia Tech, Blacksburg, Virginia 24061, USA}

\author{Double Chooz Collaboration}
\author[aa]{\\Y.~Abe,}
\author[e]{J.C.~dos Anjos,}
\author[n]{J.C.~Barriere,}
\author[v]{E.~Baussan,}
\author[a]{I.~Bekman,}
\author[i]{M.~Bergevin,}
\author[y]{T.J.C.~Bezerra,}
\author[m]{L.~Bezrukov,}
\author[f]{E.~Blucher,}
% ...

\affiliation[a]{\Aachen}
% ...
\affiliation[e]{\CBPF}
\affiliation[f]{\Chicago}
% ...
\affiliation[i]{\Davis}
% ...
\affiliation[m]{\INR}
\affiliation[n]{\CEA}
% ...
\affiliation[v]{\IPHC}
% ...
\affiliation[y]{\TohokuUni}
% ...
\affiliation[aa]{\TokyoInst}
% ...

\emailAdd{cecile.jollet@iphc.cnrs.fr}
\emailAdd{anselmo.meregaglia@iphc.cnrs.fr}

\abstract{
 The Double Chooz experiment measures the neutrino mixing angle $\theta_{13}$ by detecting reactor  $\bar{\nu}_e$ via inverse beta decay. The positron-neutron space and time coincidence allows for a sizable background rejection, nonetheless liquid scintillator detectors would  profit from a positron/electron discrimination, if feasible in large detector, to suppress the remaining background.
 Standard particle identification, based on particle dependent time profile of photon emission in liquid scintillator, can not be used given the identical mass of the two particles.
 However, the positron annihilation is sometimes delayed by the ortho-positronium (o-Ps) metastable state formation, which induces a pulse shape distortion that could be used for positron identification.
 In this paper we report on the first observation of positronium formation in a large liquid scintillator detector 
 based on pulse shape analysis of single events.
 The o-Ps formation fraction and its lifetime were measured, finding the values of 44~$\%$ $\pm$ 12~$\%$ (sys.) $\pm$ 5~$\%$ (stat.) and $3.68$~ns $\pm$ 0.17~ns (sys.) $\pm$ 0.15~ns (stat.) respectively, in agreement with the results obtained with a dedicated positron annihilation lifetime spectroscopy setup. 
}

\begin{document} 

\maketitle
\flushbottom


\section{Introduction}

Recent results of anti-neutrino reactor experiments, i.e. Double Chooz~\cite{Abe:2014bwa,Abe:2013sxa,Abe:2014lus},  Daya Bay~\cite{An:2013zwz} and RENO~\cite{Ahn:2012nd}, clearly proved the non-zero value of the $\theta_{13}$ mixing angle. 
Anti-neutrino detection, based on the inverse beta decay (IBD) process ({\it i.e. $\bar{\nu}_e + p \to e^+ + n$}), allowed a measurement of $\sin^2(2\theta_{13})$ of $\sim 0.1$.

The space and time correlation between the prompt signal given by the positron, and the delayed one given by the neutron absorption on hydrogen or gadolinium, allows for a clear signal signature and strong background reduction. Nonetheless correlated events due to fast neutrons or cosmogenic generated radio-nuclides such as $^9$Li or $^8$He still remain as a background.

A technique used in liquid scintillator detectors to extract the signal from background is pulse shape discrimination (PSD) (see Ref.~\cite{Ranucci:1998bc} and references therein). Different particles have a different energy loss while crossing the scintillator media, resulting in a particle-dependent  time profile of photon emission.
This technique is quite effective in separating light particles (e.g.\ electrons or positrons) from heavy ones (e.g. protons or alphas), however it is not adequate to distinguish between particles with similar energy losses such as positrons and electrons.

An alternative PSD, based on the observation of the ortho-positronium (o-Ps) was proposed~\cite{Franco:2010rs} for $\beta^+/\beta^-$ discrimination.
The positron emitted in the IBD process annihilates with an electron in matter, sometimes forming a positronium metastable state, which leads to a delayed annihilation.
o-Ps is the positronium triplet state which decays into three $\gamma$'s with a lifetime in vacuum of 142~ns. In matter, however, the o-Ps lifetime can be quenched by several factors, such as chemical reactions (oxidation or compound formation), magnetic effects (spin--flip), or by positronium interactions with the surrounding electrons (pick--off), yielding a two-$\gamma$ decay~\cite{PsBook}. Positronium can also be formed as a singlet state, which decays into two $\gamma$'s, called para-positronium (p-Ps), however its lifetime of $\sim 125$~ps is too short to exploit it for particle discrimination.
The o-Ps lifetime has been measured in the most commonly used solvents for organic liquid scintillators in neutrino physics~\cite{Franco:2010rs, Kino:2000}, and as a function of the dopants concentration~\cite{Consolati:2013rka}.
Although the o-Ps lifetime is typically shortened down to a few nanoseconds, the distortion in the photon emission time distribution is still observable. The Borexino collaboration~\cite{Collaboration:2011nga} has exploited the signature provided by the o-Ps induced pulse shape distortion to statistically identify and reject cosmogenic $^{11}$C $\beta^+$ decays, the dominant  background in the solar {\it pep} neutrino rate measurement.

In this paper we report the observation of o-Ps in Double Chooz, which is performed, for the first time in a large liquid scintillator detector, on pulse shape analysis of single events. The identification of the o-Ps could be used in the future as additional handle in the signal selection, reducing the cosmogenic background due to $^{9}$Li or uncorrelated background due to accidentals ($\beta^+ + n$ vs. $\beta^- + n$ chain).
Our results fully demonstrate the proof of principle and the capability of this technique, which could be a key point for the development of scintillators for future neutrino experiments.

The analysis is performed on the data set published in Ref.~\cite{Abe:2012tg}, corresponding to a live-time of about 228 days.\\

\section{Ortho-positronium properties in Double Chooz scintillators}
 \label{sec:NuToPs}

For the Double Chooz detector~\cite{Abe:2012tg} 
two different scintillators are used: one for the Target and one for the Gamma Catcher.
The liquid scintillator used for the Target is a mixture of n-dodecane, PXE, PPO, bis-MSB and 1~g gadolinium/l as a beta-diketonate complex. The Target volume is surrounded by the Gamma Catcher scintillator, which is similar to the Target (mineral oil, n-dodecane, PXE, PPO, bis-MSB) but Gd-free~\cite{Aberle:2011ar,Aberle:2011zm}. 
The light yield of the Gamma Catcher was chosen to provide identical photoelectron yield across these two layers.

To measure the formation fraction of o-Ps and its lifetime in samples of the Double Chooz scintillators, a standard PALS (Positron Annihilation Lifetime Spectroscopy) system made of two plastic scintillator (BaF$_2$) detectors has been used. 
The apparatus, at IPHC Strasbourg, is very similar to the one described in detail in Ref.~\cite{Consolati:2013rka}. 
The positron source is a 1~MBq $^{22}$Na source, inserted between two 12.7~$\mu$m titanium layers, immersed in the liquid scintillator. The 1.27~MeV $\gamma$-ray emitted in association with the positron is detected by one detector (lower threshold at 950~keV) and used as trigger, whereas the second detector (400--700~keV energy range) is dedicated to the measurement of one $\gamma$-ray of 511~keV coming from the positron annihilation or one from the o-Ps decay; the time between the two signals is measured to reconstruct the o-Ps lifetime. The overall time resolution of the apparatus is $\sim 180$~ps.

\begin{figure} [phtb]
\begin{center}
\includegraphics[width=0.63\columnwidth]{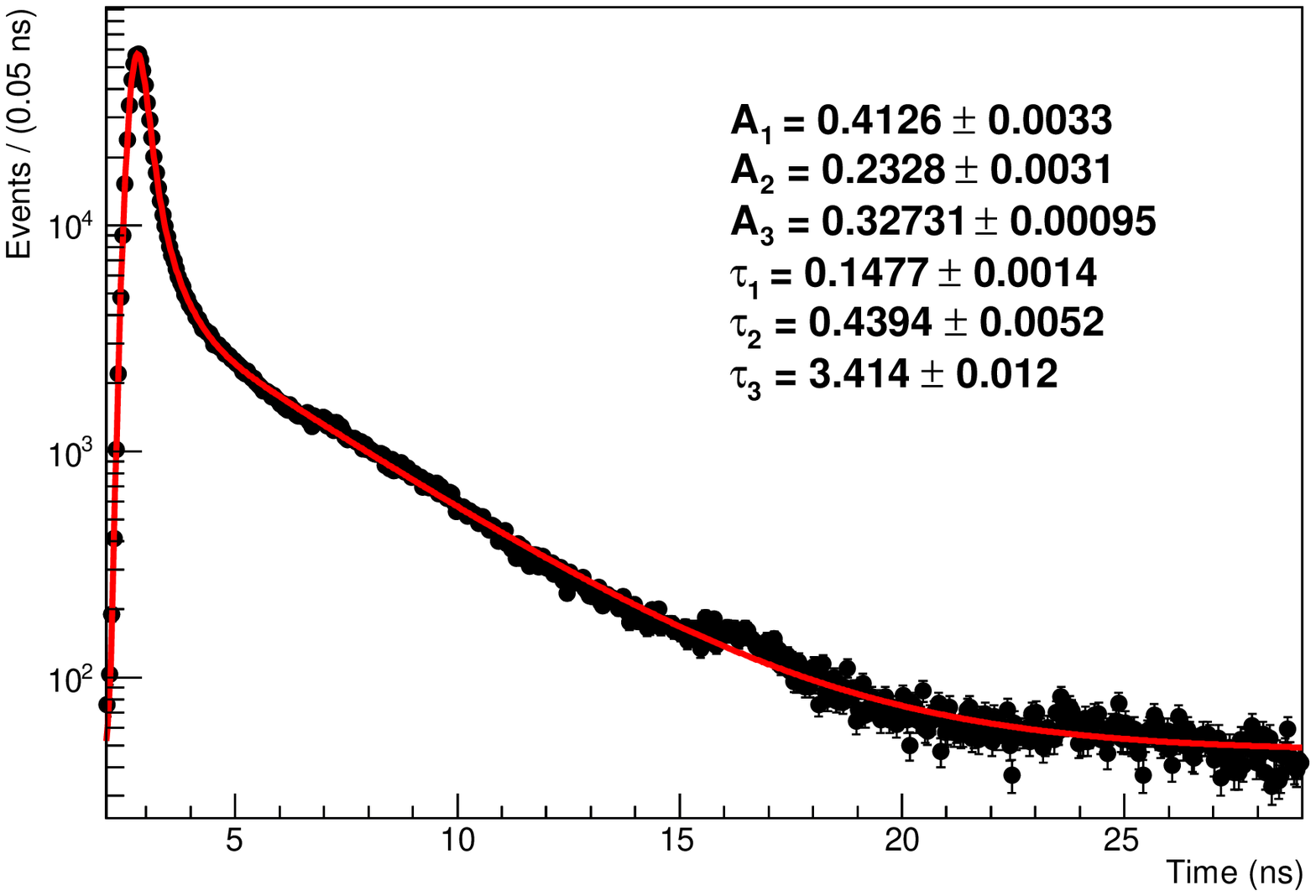}
\includegraphics[width=0.63\columnwidth]{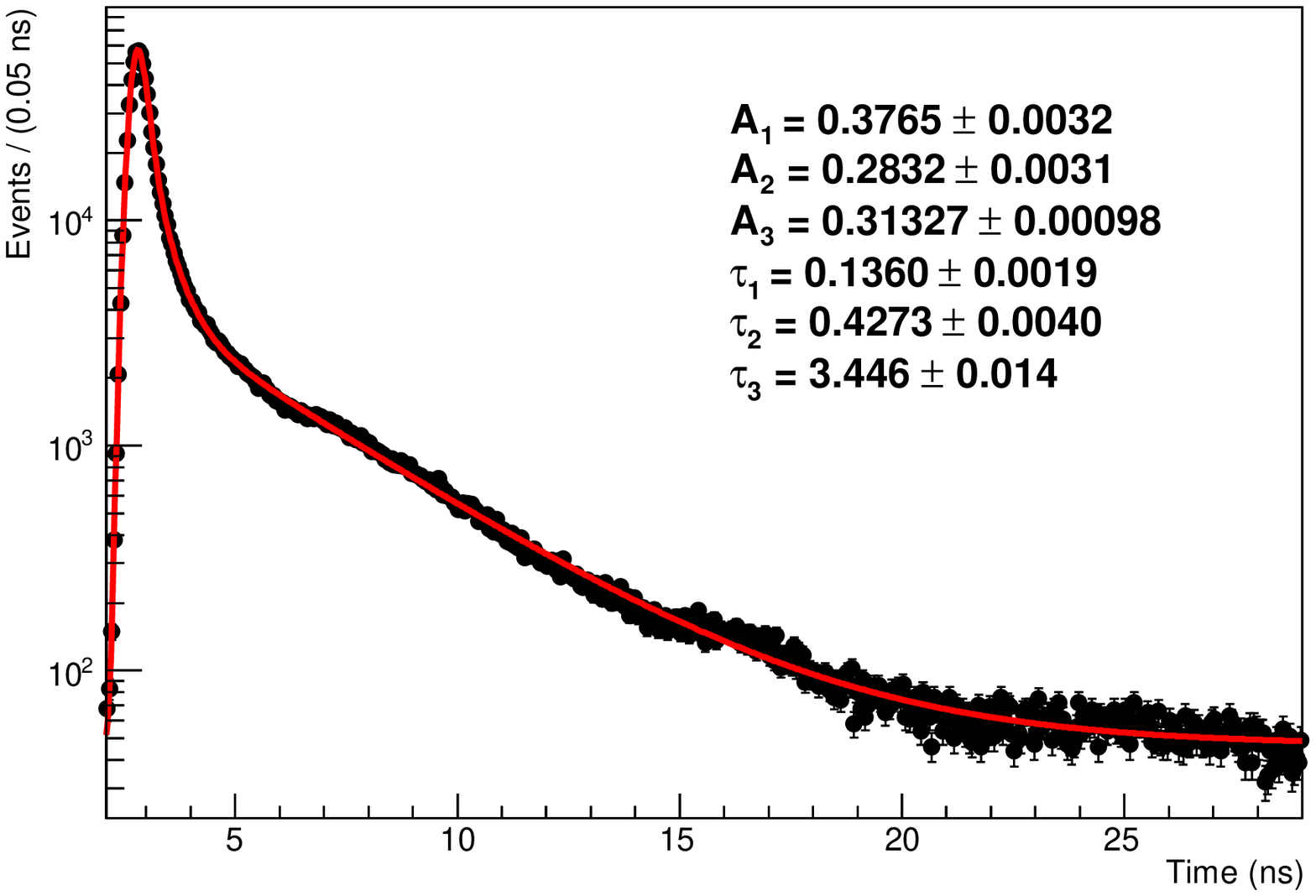}
\caption{{\it Time distribution (black dots) and corresponding fit (red line) for the Target (top) and Gamma Catcher (bottom) scintillator.}}
\label{fig:oPsT}
\end{center}
\end{figure}

The measured time distribution is fitted with a combination of three exponentials ($A_i \cdot e^{-t/\tau_i}$) and a constant $C$, all convoluted with a Gaussian spread to model the detector resolution:
\begin{equation}
F(t)= \sum\limits_{i=1}^3 A_i\cdot e^{-t/\tau_i} + C. 
\end{equation}
 The two short exponentials correspond to the positron annihilation and p-Ps formation and decay in the source structure and in the liquid, whereas the long one corresponds to the o-Ps lifetime. The constant function is used to take into account pile-up and random noise events. The details of the analysis can be found in Ref.~\cite{Consolati:2013rka}.
The obtained time distributions and the corresponding fits are shown in Fig.~\ref{fig:oPsT} for Target and Gamma Catcher scintillator. The lifetime can be taken immediately from the fit parameter $\tau_3$ whereas the o-Ps formation fraction has to be computed renormalizing correctly the number of events, neglecting the events in the titanium source support, and taking into account the different efficiencies for the three and two-$\gamma$ decay modes.\\

Once the systematics are considered (they account typically for about 1.7~\% on the formation fraction and for about 1~\% on the lifetime~\cite{Consolati:2013rka}) we obtain for the Gamma Catcher scintillator an o-Ps fraction formation of $45.6 \pm 1.3$~\% and lifetime of $3.45 \pm 0.03$~ns.
Slightly different values were found for the Target: a formation fraction of $47.6 \pm 1.3$~\% and a lifetime of $3.42 \pm 0.03$~ns. All the results are summarized in Tab.~\ref{tab:resnutops}.

Although Target and Gamma Catcher yield values compatible within the errors for both o-Ps formation and lifetime, high precision measurements on different liquid scintillators showed that o-Ps formation has a clear dependence on the scintillator loading whereas the lifetime is much more stable~\cite{Consolati:2013rka}.

\begin{table}[t]
\begin{center}
\begin{tabular}{|c|c|c|}
\hline
Scintillator &  o-Ps formation fraction & o-Ps lifetime\\
\hline
Target & 47.6 $\pm$ 1.3 \% & 3.42 $\pm$ 0.03 ns\\
Gamma Catcher & 45.6 $\pm$ 1.3 \% & 3.45 $\pm$ 0.03 ns\\
\hline
\end{tabular}
\caption{{\it o-Ps formation fraction and lifetime in Double Chooz scintillator of Target and Gamma Catcher measured with a dedicated PALS setup~\cite{Consolati:2013rka}.}}
\label{tab:resnutops}
\end{center}

\end{table}%

\section{Pulse shape reconstruction in Double Chooz}

\begin{figure} [htbp]
\begin{center}
\includegraphics[width=0.63\columnwidth]{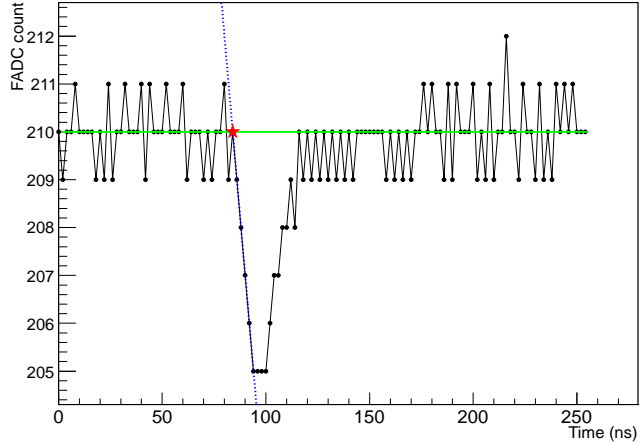}
\caption{{\it Example of the time determination of the pulse for one PMT. The green solid line represents the pedestal value (FADC=210), the blue dashed line shows the fit of the pulse edge. The pulse starting time (84 ns) is represented by a red star.}}
\label{fig:TimePMT}
\end{center}
\end{figure}

In Double Chooz,  the scintillation signal is recorded by 390 10-inch PMTs, installed on the inner wall of the stainless steel buffer tank (see Ref.~\cite{Abe:2012tg} for details).
At each trigger a waveform of 256~ns is recorded for each PMT using flash-ADC (FADC)~\cite{Abe:2013iha}. 
The PMT baseline is computed using the first thousands events of the run, selecting the ones in which the PMT itself did not record any signal above threshold (the threshold was set to 2 FADC counts which corresponds to approximately 0.3~photo-electrons). When a signal is observed on a PMT, a linear fit is performed on its rising edge (see Fig.~\ref{fig:TimePMT}). The intercept of the fit line with the baseline provides the pulse starting time and represents one entry of the time profile distribution. The obtained event time profile is the distribution of the arrival time of the pulses recorded by all PMTs.

To correctly build the time profile distribution, the time of flight between the reconstructed vertex position and the PMT is subtracted for each pulse. In addition, the calibration of time offset for each channel, as measured with a laser and monitored with a LED system, is accounted for.

Note that the same PMT can of course record more than one pulse in the same event. The minimum time between two pulses on a single PMT was set to 25~ns (i.e. all that happens within this time window is merged into a single pulse), however tests with a narrower time window set to 10~ns showed no significant influence on the result on this parameter.

Once all the time pulses in one event are computed, they are sorted and shifted in order to have the first one equal to zero. This shift is needed to correctly compare time profile of different events.

\begin{figure} [htbp]
\begin{center}
\includegraphics[width=0.63\columnwidth]{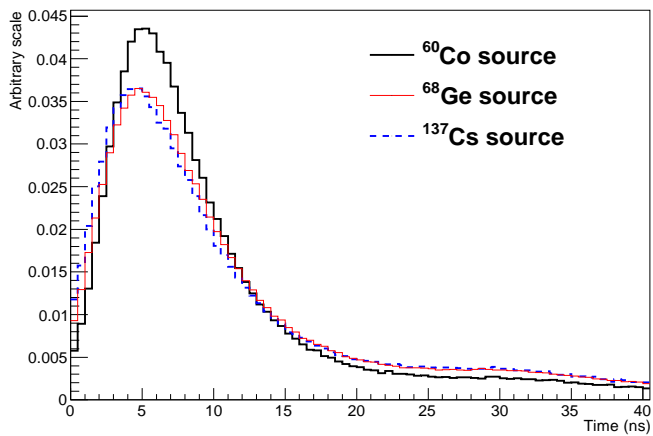}
\caption{{\it Time profile distribution for $^{60}$Co (thick black line), $^{137}$Cs (dashed blue line) and $^{68}$Ge (thin red line) events in the center of the Target. The distributions are normalized to one.}}
\label{fig:CumulativeCobalt+Cesium}
\end{center}
\end{figure}

This procedure was carried out on $^{60}$Co, $^{137}$Cs and $^{68}$Ge calibration runs, with the sources located at the center of the detector, and the resulting time profile of all the events are shown in Fig.~\ref{fig:CumulativeCobalt+Cesium}. 
The small difference between the obtained time profiles can be interpreted as a slight energy dependence. This was confirmed looking at the changes of the reconstructed time profile for neutrino candidates when computed in different ranges of visible energy.\\

\section{Ortho-positronium tagging algorithm}
\label{sec:algo}

All the prompt signals of IBD events are composed of an e$^{+}$ ionization signal, followed by the two 511~keV-$\gamma$ rays emission. If o-Ps is not formed, the time between these two processes is too short to distinguish them, however, in case of o-Ps formation the delay between the two signals is no longer negligible since o-Ps has a lifetime of about 3.4~ns in the Double Chooz liquid scintillators.

The discrimination between ionization and annihilation signals becomes in principle possible, but it is not trivial given the scintillator fast decay time of 2.6~ns in the Target and  5.4~ns in the Gamma Catcher (GC)~\cite{Aberle:2011ar}. In addition, faster electronics than used
which has a time sample width of 2~ns, would be desirable to provide a better time resolution.

The idea is to look for a double peak pattern in the time profile: the first one is due to positron ionization, and the second one to the emission of the two 511~keV-$\gamma$ rays.
Such a pattern can be better observed for events in which the o-Ps has a long lifetime (large time interval between the peaks on the time profile distribution) and for low energy events (second peak less hidden by the tail of the first signal).

Based on this idea a specific algorithm was developed: a fit function was built combining two reference time profiles separated by a delay corresponding to the time taken by o-Ps to decay.
The parameters considered and their allowed range are quoted here below.

\begin{itemize}
\item $\Delta t$: time interval between the two reference time profiles. It is allowed to vary between 0 (no o-Ps formation observed) and the shortest time for which the number of pulses in the following 50~ns is lower than $70~\%$ of the number of pulses corresponding to the 1.022~MeV signal.\\

\item $\epsilon_1$, $\epsilon_2$: normalization of the two reference time profiles. The normalization evaluation is based  on the visible energy: the second peak corresponds to a signal of 1.022~MeV and the first one to the remaining energy. The relative normalization is computed assuming that the number of pulses is proportional to the visible energy. This is not completely correct, in particular as the energy increases, since a pulse can correspond to more than a single photo-electron. To overcome this issue, each normalization was allowed to vary in a range of $\Delta_\epsilon = \pm 0.6 \, \epsilon$, and the uncertainty $\sigma$ of $0.2~\epsilon $ was used in the $\chi^2$ computation.
The number of multiple photo-electrons pulses could be estimated from the visible energy reducing the error on the normalization, however this would rely on the vertex position introducing additional systematics related to the vertex reconstruction.\\

\item $\lambda$: shift of the first reference time profile. Since not all the event time profiles start exactly at zero, a possible shift is taken into account in a range between $-10$~ns and the time of the first pulse in the event time profile.\\
\end{itemize}

If $f_{\rm TP}$ is the function representing the reference time profile, each event time profile is therefore fitted using the following function $f_{\rm fit}$:

\begin{equation}
f_{\rm fit} (t)= (\epsilon_1+ \Delta_{\epsilon 1}) f_{\rm TP}(t - \lambda) + (\epsilon_2+ \Delta_{\epsilon 2}) f_{\rm TP}(t - \lambda - \Delta t).
\end{equation}

\begin{figure} [phtb]
\begin{center}
\includegraphics[width=0.62\columnwidth]{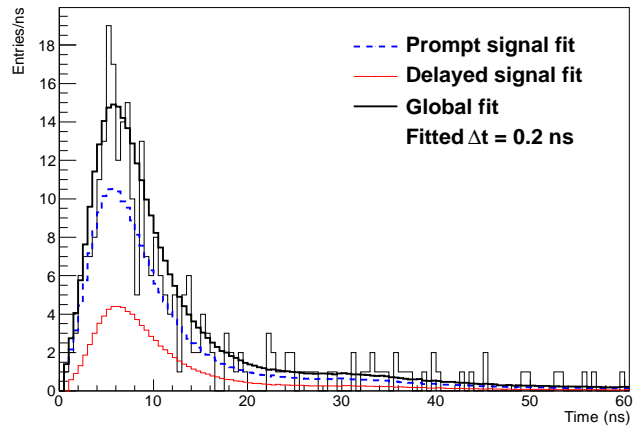}
\includegraphics[width=0.62\columnwidth]{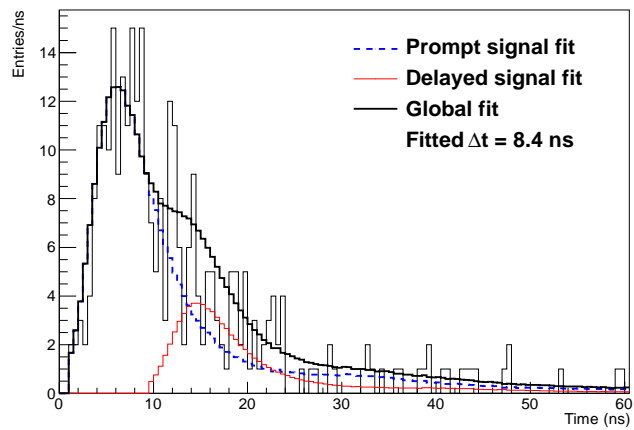}
\includegraphics[width=0.62\columnwidth]{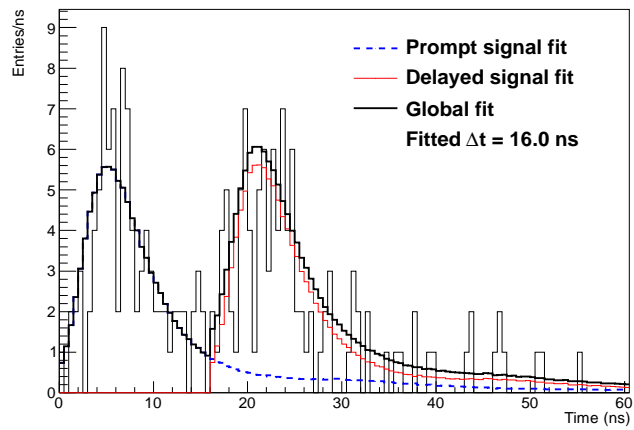}
\caption{{\it Three examples of o-Ps fit. The dashed blue line represents the fit of the first time profile, the thin red line the fit of the second one and the thick black line is the total fit.}}
\label{FitDeltaT}
\end{center}
\end{figure}

The number of pulses per event is of the order of 400, therefore the number of entries per bin (300 bins of 0.5~ns) is rather small and it is Poisson distributed. 
The following $\chi^2$ definition~\cite{Beringer:1900zz} was therefore used: 
\begin{equation}
\chi^2= 2 \sum_{i=1}^{N} \left \lbrack \nu_i-n_i+n_i\, ln \frac{n_i}{\nu_i} \right \rbrack + \sum_{j=1}^{2}\frac{\Delta_{\epsilon j}^2}{\sigma_{j}^2}
\end{equation}
where $N$ is the number of bins (i.e. 300), $n_i$ and $\nu_i$ are the number of pulses observed and expected in the $i^{\rm th}$ bin respectively, and only bins for which $n_i > 0$ are considered.
The term $\Delta_{\epsilon j}$ with $j=1,2$ represents the normalization variations of the two reference time profiles with respect to the computed ones, and $\sigma_j$ the 20~\% normalization error.
The $\chi^2$ is computed using the MINUIT minimization embedded in the ROOT package~\cite{Brun:1997}.

To discard from the analysis events for which the fit did not converge properly, an upper limit of 2 was applied on the reduced $\chi^2$. For the same reason, an additional constraint was applied on the ratio between the integral of the fitted function and the number of pulses, in the first and second peak region separately. Events with an integral ratio larger than 2 in either the first or the second peak region (i.e. the interval between 3~ns before the peak and 10~ns after it) are discarded.
The applied cuts removed 0.2~\% of the signal and resulted in no bias in the event selection.

As examples, the result of the fit obtained for events with $\Delta t$ of 0.2~ns, 8.4~ns and 16.0~ns are shown in Fig.~\ref{FitDeltaT}.

\section{Results on the ortho-positronium properties}
\label{sec:analysis}

To unambiguously demonstrate the capability of Double Chooz to observe the o-Ps formation, the algorithm is applied to a pure $^{60}$Co sample and compared to the results obtained for the neutrino candidates.
The $^{60}$Co reference time profile obtained locating the radioactive source at the center of the detector was used for this analysis. Systematic uncertainty related to the choice of reference time profile will be evaluated later on in this section.

Since an increase of the error of the reconstructed vertex is known to rapidly degrade the time profile, a maximal distance of 20~cm between the position where the source was deployed and the reconstructed vertex was required for a clean sample selection. Such a selection could only be applied on the source calibration data used to build the reference time profile.

As far as the neutrino candidate selection is concerned, an additional requirement on the energy with respect to the selection cut of Ref.~\cite{Abe:2012tg} is applied: only events with a visible energy between 1.2 and 3~MeV are analyzed. Below 1.2~MeV the first peak energy is too small (i.e. below 200~keV) for a correct fit convergence, whereas above 3~MeV the second peak is typically hidden by the first peak's tail and therefore difficult to identify.

\begin{figure} [htbp]
\begin{center}
\includegraphics[width=0.63\columnwidth]{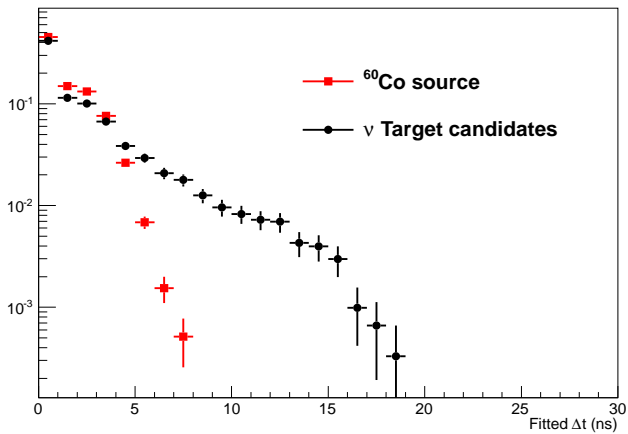}
\caption{{\it Distribution of the $\Delta t$ value determined by the fit for the cobalt sample (red squares), and for the neutrino sample (black dots), normalized to one.}}
\label{fig:DeltaTNeutrinos+Cobalt}
\end{center}
\end{figure}

 As can be seen in Fig.~\ref{fig:DeltaTNeutrinos+Cobalt} (red squares), the $\Delta t$ distribution for the $^{60}$Co events is, as expected, peaked at zero. Nonetheless, a smearing can be seen resulting in $\Delta t$ values up to 7~ns. This smearing shows the present limitation of the developed algorithm, which sometimes tends to increase the $\Delta t$ parameter reducing the value of the shift $\lambda$ of the first reference time profile. Indeed the fluctuations of the tail of a time profile can sometimes be wrongly identified as a second signal.
 
In the same figure (Fig.~\ref{fig:DeltaTNeutrinos+Cobalt}) the distribution obtained for the neutrino selection of the gadolinium analysis of Ref.~\cite{Abe:2012tg} is shown (black dots) for a direct comparison between the two.  A clear excess of events at large $\Delta t$ is present in the neutrino sample with respect to the cobalt one. This shows indeed the Double Chooz capability to observe o-Ps formation.

\begin{figure} [phtb]
\begin{center}
\includegraphics[width=0.63\columnwidth]{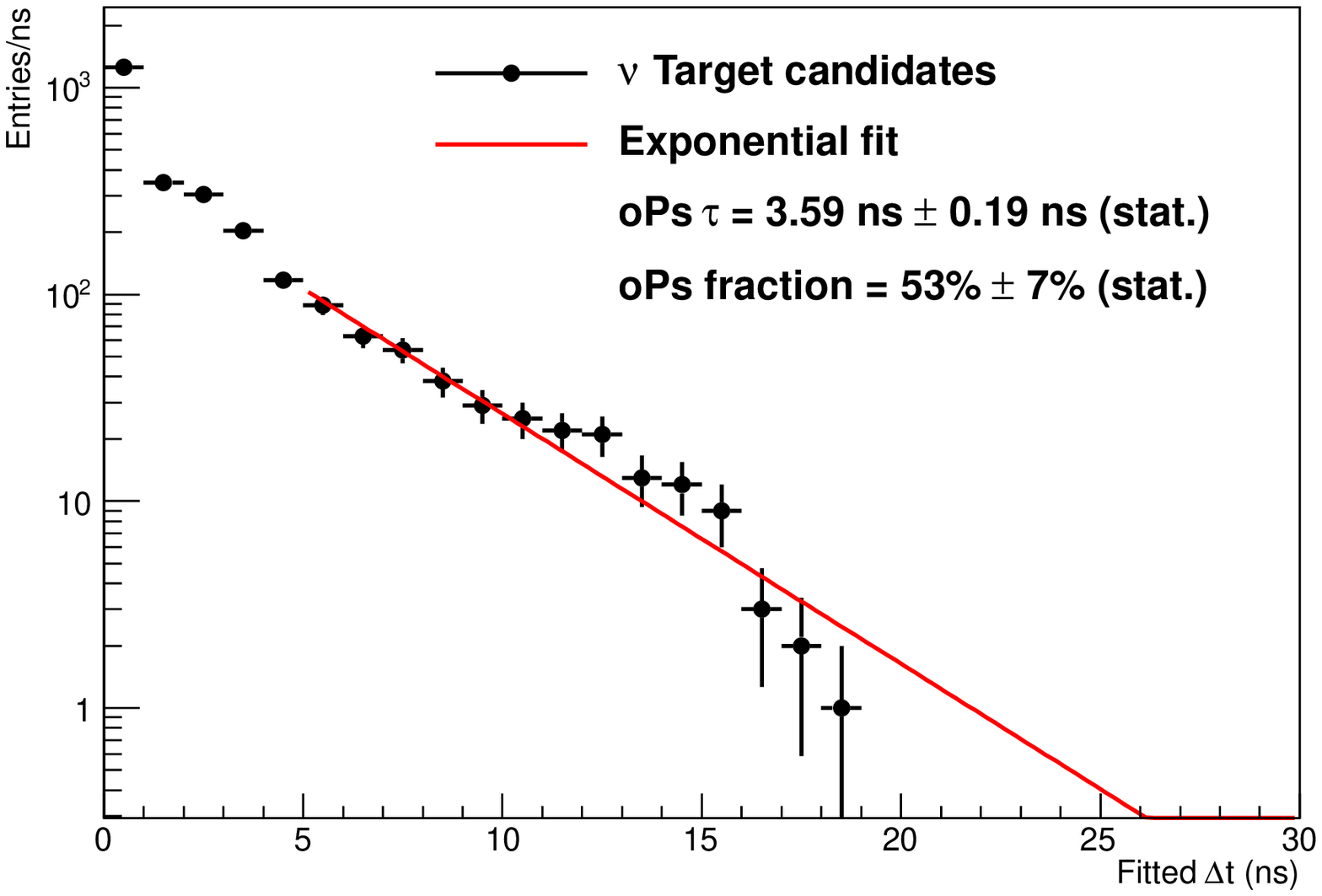}
\includegraphics[width=0.63\columnwidth]{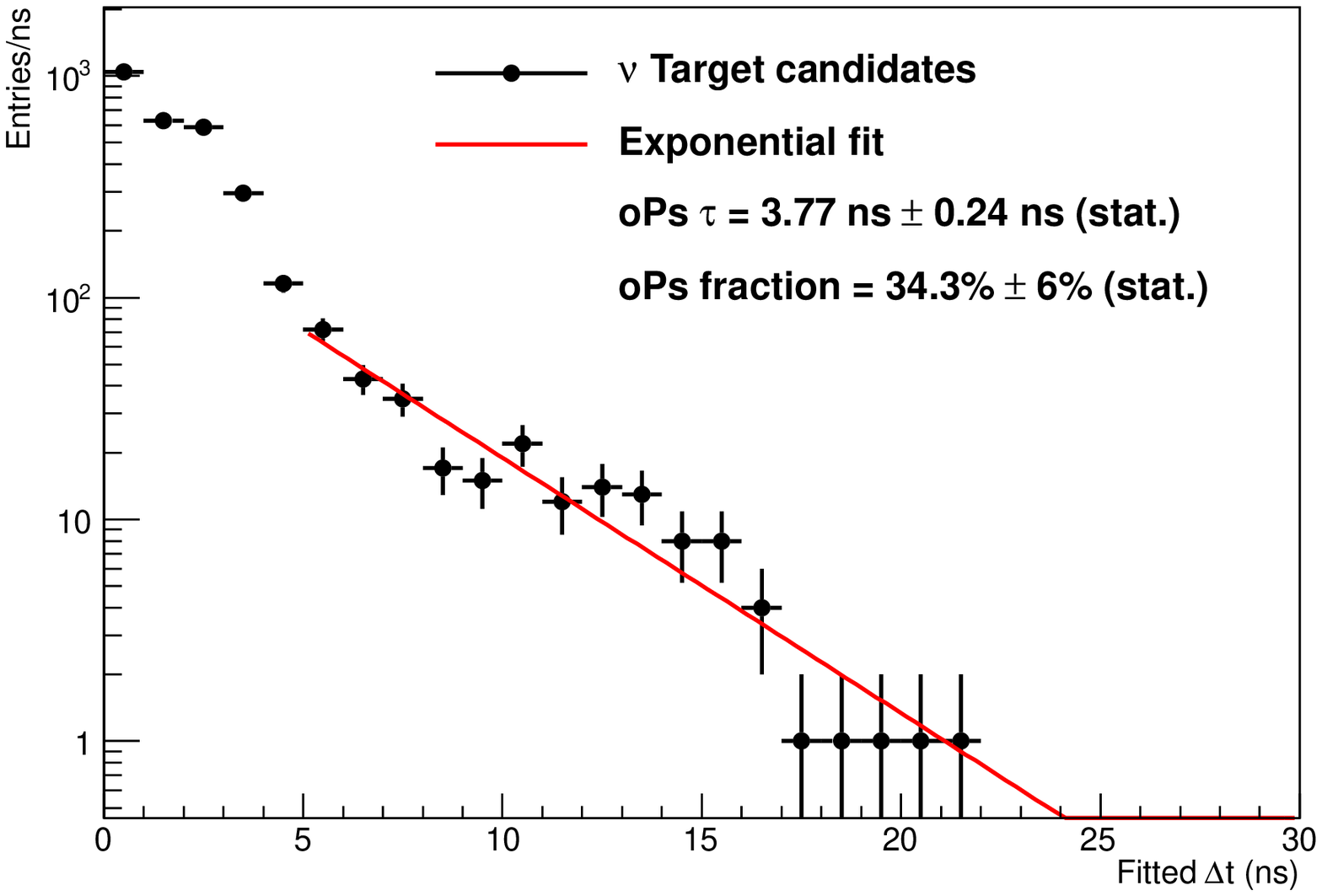}
\caption{{\it Fit of the $\Delta t$ distribution, for neutrino candidates with energy between 1.2 and 3~MeV, with an exponential function for $\Delta t$ greater than 5 ns. The $^{60}$Co (top) and the $^{137}$Cs (bottom) reference time profiles are used in the oPs tagging algorithm. The errors quoted on the figures are statistical only.}}
\label{fig:LE}
\end{center}
\end{figure}

To estimate the lifetime and the o-Ps production fraction, the obtained $\Delta t$ distribution is fitted with an exponential function with a lower bound of 5~ns, since this corresponds to the end of the smearing of the zero $\Delta t$ events as observed in the $^{60}$Co sample study. A higher value for the fit lower bound could be chosen granting a higher purity o-Ps sample, however the reduced statistics would lead to a larger error. Tests were made using values between 4 and 8~ns for the lower boundaries and the results were used to compute systematics related errors.
Moreover, the algorithm outcome has a dependence on the choice of the reference time profile and this has also to be evaluated as systematic uncertainties.
Additional parameters having an impact on the final results are the vertex distance cut applied when building the reference time profile for a given calibration source, and its position inside the detector.
To study all these aspects the analysis of the neutrino candidates was carried out under different conditions, namely:
\begin{enumerate}
\item Using as reference time profile the one obtained with a $^{60}$Co and a $^{137}$Cs source.
\item Modifying the distance cut between the reconstructed vertex and the true source position between 10 and 30~cm when building the reference time profile.
\item Selecting different positions inside the detector of the same source to build the reference time profile.
\item Changing the lower bound of time interval of the fit between 4 and 8~ns.
\end{enumerate}
Whereas for items 2 to 4 it is reasonable to assume that the best estimate is the mean of the results and to take as systematic uncertainty the spread, i.e. the r.m.s., of the different measurements, for item number 1 the situation is different.
The best estimate corresponds to the measurement obtained with the reference time profile which has the energy that best matches the one of the neutrino candidate events.

For the neutrino candidates sample, the mean visible energy in the selected region from 1.2 to 3~MeV, is 2.237~MeV, which is more similar to the high energy bound given by the $^{60}$Co (2.5~MeV) rather than the low energy one of $^{137}$Cs (0.66~MeV). However, if o-Ps is formed we have two separate signals of 1.215~MeV (mean positron prompt energy) and 1.022~MeV (o-Ps decay) respectively, therefore the best reference time profile would be in between the two energy regimes.

The systematics due to the choice of the source for the reference time profile ($\sigma_{\rm Ref}$) is therefore evaluated as half the difference between the low energy (i.e. $^{137}$Cs) and high energy (i.e. $^{60}$Co) regimes:
\begin{equation}
\sigma_{\rm Ref} = \frac{1}{2} (\text{V}_{\text{Co}} - \text{V}_{\text{Cs}})
\end{equation}
where V$_{\text{X}}$ is the value (i.e. either the o-Ps lifetime or its formation fraction) measured for the reference time profile obtained with the source X.

The fits obtained using the $^{60}$Co and $^{137}$Cs reference time profiles are shown in Fig.~\ref{fig:LE}, together with the results and the statistical errors, whereas a summary of all the different contributions to the systematic error is given in Tab.~\ref{tab:EvalSystTarget}.
Note that in the fraction formation evaluation, the presence of background events in the neutrino candidate sample is accounted for. The expected background, in the 1.2 to 3~MeV energy region, is $\sim 0.5$ events per day, whereas the neutrino candidates rate in the same region, after the applied analysis cuts reported at the end of Sec.~\ref{sec:algo}, is $\sim 13.3$ events per day. The correction due to the presence of background corresponds therefore to about 3.8~\%.

Adding quadratically the four systematic contributions, and taking the mean between the o-Ps properties measured using $^{137}$Cs and $^{60}$Co, it can be stated that the formation fraction and the lifetime in the Target scintillator were observed to be 44~\% $\pm$ 12~\% (sys.) $\pm$ 5~\% (stat.) and 3.68~ns $\pm$ 0.17~ns (sys.) $\pm$ 0.15~ns (stat.) respectively, in good agreement with the expectations. \\

\begin{table}[t]
\begin{center}
\begin{tabular}{|c|c|c|}
\hline
Error   & o-Ps formation  & o-Ps lifetime \\
  type & fraction error [$\%$]&  error [ns]\\

\hline
Source element  &  9  & 0.09 \\
Cut on the vertex distance & 1.25  & 0.019 \\
Source position & 5  & 0.055 \\
Fit interval & 7  & 0.14 \\
\hline
Total systematics & 12 & 0.17\\
\hline
Statistics & 5  & 0.15 \\
\hline
\end{tabular}
\caption{{\it Evaluation of the different statistical and systematic uncertainties.}}
\label{tab:EvalSystTarget}
\end{center}
\end{table}

\section{Possible physics impact}

The Borexino experiment has already proven the possibility to use o-Ps observation on a statistical way for $e^+/e^-$ discrimination reducing the cosmogenic $^{11}$C background in the solar $pep$ neutron observation~\cite{Collaboration:2011nga}.
The demonstrated capability of Double Chooz to observe o-Ps formation on pulse shape analysis of single events is an important step forward in this technique, that could be exploited for the neutrino event selection and background reduction in liquid scintillation detectors. 

In experimental projects aiming at the electron antineutrino detection, the $e^+/e^-$ discrimination could be directly used to reduce the background.
This is not the case of IBD based experiments. Since o-Ps is formed only in about half of the events, a selection cut based on o-Ps observation is not conceivable for antineutrino events selection in IBD based experiments, even assuming a perfect detection efficiency.
In this case, however, the o-Ps detection could be used to select a clean antineutrino subsample to validate experimental results.
For example, one of the largest background contribution in IBD experiments, comes from the cosmogenic $^9$Li events, for which the prompt signal
is an electron resulting therefore in no o-Ps formation observation.

In order to profit the most from the o-Ps signature, several improvements would be desirable, in particular the liquid scintillator should be optimized to have a fast scintillating time and the longest possible o-Ps lifetime. Another critical parameter is the detector light yield: a large light yield would provide an easier o-Ps observation over a broader energy range. In addition, further improvements could come from the detector electronics, since a fast
(sub-nanosecond) readout electronics would allow for the observation of small pulse shape distortions.

\section{Conclusions}

o-Ps formation could be exploited in anti-neutrino detector for additional background rejection.
Although its detection is quite challenging in large liquid scintillator detectors due to its short lifetime of about 3~ns, this process has already been used in neutrino physics for a $e^+/e^-$ discrimination~\cite{Collaboration:2011nga} on a statistical basis.

Double Chooz has demonstrated for the first time the possibility to tag such a process on pulse shape analysis of single events. Relying on a selection algorithm based on the pulse shape distortion of the event, o-Ps formation was observed and its lifetime and formation fraction were measured, finding a good agreement with the values obtained in laboratory measurements.

The obtained result is so far energy dependent and it can not be used directly for a background reduction in Double Chooz, however it is now possible to assign a probability of each event of being an o-Ps decay for dedicated studies on pure samples.

Considering that the Double Chooz detector was not conceived for such a measurement (both in the choice of electronics and scintillator), this result is also an excellent starting point for future projects aiming at the liquid scintillation technology for anti-neutrino detection.

\acknowledgments
We thank the French electricity company EDF; the European fund FEDER; the R\'egion de Champagne Ardenne; the D\'epartement des Ardennes; and the Communaut\'e des Communes Ardennes Rives de Meuse. We acknowledge the support of the ANR NuToPs project (grant 2011-JS04-009-01), the CEA, CNRS/IN2P3, the computer center CCIN2P3, and
the UnivEarthS LabEx program of Sorbonne Paris Cit\'e (ANR-10-LABX-0023 and ANR-11-IDEX-0005-02) in France; the Ministry of Education, Culture, Sports, Science and Technology of Japan (MEXT) and the Japan Society for the Promotion of Science (JSPS); the Department of Energy and the National Science Foundation of the United States; the Ministerio de Ciencia e Innovaci\'on (MICINN) of Spain; the Max Planck Gesellschaft, and the Deutsche Forschungsgemeinschaft DFG (SBH WI 2152), the Transregional Collaborative Research Center TR27, the excellence cluster ``Origin and Structure of the Universe'', and the Maier-Leibnitz-Laboratorium Garching in Germany; the Russian Academy of Science, the Kurchatov Institute and RFBR (the Russian Foundation for Basic Research); the Brazilian Ministry of Science, Technology and Innovation (MCTI), the Financiadora de Estudos e Projetos (FINEP), the Conselho Nacional de Desenvolvimento Cient\'{\i}fico e Tecnol\'ogico (CNPq), the S\~ao Paulo Research Foundation (FAPESP), and the Brazilian Network for High Energy Physics (RENAFAE) in Brazil.

\end{document}